\documentclass[reprint,amsmath,amssymb,aps,prd]{revtex4-2}

\usepackage{amsmath,amssymb,latexsym,slashed,multirow}

\usepackage{graphicx}

\def\lmatrix{\left(\begin{array}}
\def\rmatrix{\end{array}\right)}
\def\bea{\begin{eqnarray}}
\def\eea{\end{eqnarray}}
\def\nn{\nonumber}
\def\msbar{\overline{\rm MS\kern-0.5pt}\kern0.5pt}
\def\gsim{\mathrel{\rlap{\lower4pt\hbox{\hskip1pt$\sim$}}\raise1pt\hbox{$>$}}}
\def\lsim{\mathrel{\rlap{\lower4pt\hbox{\hskip1pt$\sim$}}\raise1pt\hbox{$<$}}}
\def\rho{\varrho}
\def\eps{\varepsilon}

\begin{document}

\title{The $f_\rho / m_\rho$ and $f_\pi / m_\rho$ ratios and the conformal window}

\author{Hee Sok Chung} 
\email{neville@korea.ac.kr}
\affiliation{Korea University, Department of Physics, Seoul 02841, Korea}

\author{Daniel Nogradi}
\email{nogradi@bodri.elte.hu}
\affiliation{Eotvos University, Institute for
  Theoretical Physics, Pazmany Peter setany 1/A, 1117, Budapest,
  Hungary}

\begin{abstract}
    The $f_\rho / m_\rho$ ratio is calculated at N$^3$LO order within perturbative (p)NRQCD with $N_f$ flavors of mass
    degenerate fermions. The massless limit of the ratio is expanded \'a la Banks-Zaks in $\eps =
    16.5 - N_f$ leading to reliable predictions close to the upper end of the conformal window.
    The comparison of the NNLO and N$^3$LO results indicate that the Banks-Zaks
    expansion may be reliable down to twelve flavors.
    Previous lattice calculations combined with the KSRF relations
    provide us with the same ratio for the range $2 \leq N_f \leq 10$. 
    Assuming a monotonous dependence on $N_f$ leads to an estimate for the lower end of the conformal
    window, $N_f^* \simeq 12$, by matching the non-perturbative and our perturbative results. In any case an
    abrupt change is observed in $f_\rho / m_\rho$ at twelve flavors.
    As a cross-check we also consider the $f_\pi / m_\rho$ ratio for which lattice results are also
    available. The perturbative calculation at present is only at the NNLO level
    which is insufficient for a reliable and robust matching between the low $N_f$ and high $N_f$
    regions. Nonetheless, using the relative size of the N$^3$LO correction of $f_\rho / m_\rho$ for
    estimating the same for $f_\pi / m_\rho$ leads to the estimate $N_f^* \simeq 13$.
\end{abstract}

\maketitle

\section{Introduction and summary}
\label{introduction}

How gauge theories with spontaneous chiral symmetry breaking 
transition into conformal gauge theories as the massless fermion content is increased 
\'a la Banks-Zaks is a non-trivial QFT problem \cite{Banks:1981nn}. We propose dimensionless
ratios of meson decay constants and masses as promising candidates to shed light on the particulars of
the transition. Concretely, we will study $f_{\rho} / m_\rho$ and $f_\pi / m_\rho$ in this paper. Our
main objective is to estimate or constrain the critical flavor number, $N_f^*$, in other words the 
lower end of the conformal window.

The non-trivial problem of finding or constraining $N_f^*$ in gauge
theories has been addressed by many different approaches in the past
\cite{Appelquist:1988yc, Cohen:1988sq, Sannino:2004qp, Dietrich:2006cm,
Armoni:2009jn, Frandsen:2010ej, Ryttov:2016ner, Ryttov:2016hdp,
Ryttov:2016hal, Kim:2020yvr, Lee:2020ihn}.  Reviews concerning the
non-perturbative lattices studies include
\cite{DeGrand:2015zxa,Nogradi:2016qek,Rummukainen:2022ekh} and
references therein. 

Clearly, a purely perturbative calculation, even at high orders, is not sufficient to determine $N_f^*$
with any degree of confidence. Some non-perturbative input is required since just below the conformal
window the theory is expected to be strongly coupled. In our work we will carry out high order perturbative
calculations valid in the high $N_f$ conformal region and combine it with non-perturbative results from the low
$N_f$ region in a meaningful way.

Below the conformal window both the nominators and denominators of our ratios have well-defined chiral limits
and are both $O(\Lambda)$, the dynamically generated scale. 
The ratios are finite and can be computed via non-perturbative lattice calculations
carefully extrapolated to the infinite volume, chiral and continuum limits. 
Inside the conformal window both decay constants and masses scale the same with the fermion mass $m$ 
and the ratios again have a well-defined chiral limit. In this way the ratios
can meaningfully be compared across the transition covering the full range of fermion content
provided asymptotic freedom is present. This observation is the main motivation for our study. The gauge
group will be $SU(3)$ throughout.

Perturbation theory is clearly not applicable below the conformal window, at low $N_f$, hence the need
for non-perturbative lattice simulations there. Continuum and chirally extrapolated lattice results are
available for $f_\pi / m_\rho$ within the range $2 \leq N_f \leq 10$ 
\cite{Nogradi:2019iek, Nogradi:2019auv, Kotov:2021mgp}. Using a KSRF-relation
\cite{Kawarabayashi:1966kd, Riazuddin:1966sw} these can be reused for $f_\rho / m_\rho$. This
non-perturbative input is essential and will supplement our perturbative results. 

Close to the upper end of the conformal window, at high $N_f$
where the fixed point coupling is small, perturbation theory is unambiguously reliable. This occurs
below $N_f = 33/2$ for $SU(3)$, the point at which asymptotic freedom is lost. In this paper
we calculate $f_\rho$ and $m_\rho$ in perturbation theory at finite fermion  mass within the framework of
(p)NRQCD to N$^3$LO accuracy, and obtain $f_\rho / m_\rho$, followed by the massless limit. 
The deviation between the NNLO and N$^3$LO
results are very small down to twelve flavors indicating convergence of the perturbative series.
Assuming $f_\rho / m_\rho$ is monotonous as a
function of $N_f$ we attempt to match the non-perturbative low $N_f$ region and the perturbative high
$N_f$ region. At twelve flavors an abrupt change occurs which we identify as an estimate of the lower end
of the conformal window, $N_f^* \simeq 12$. 

The same approach could be applied to the $f_\pi / m_\rho$ ratio as well. On the non-perturbative side
continuum and chiral extrapolated lattice results are available in the literature as already mentioned.
On the perturbative side, inside the conformal window, we are only able to calculate $f_\pi$
to NNLO order at present, one order lower than it is currently possible for $f_\rho$. Nonetheless, if we
take the relative size of the N$^3$LO result found for $f_\rho / m_\rho$ and estimate the
corresponding contribution to $f_\pi / m_\rho$ to be about the same, we may extract $N_f^*$ using the same
procedure. From $f_\pi / m_\rho$ we obtain in this way $N_f^* \simeq 13$ but of course this result should be
taken as indicative only, a genuine N$^3$LO calculation should be performed for $f_\pi$ in the future to
validate it. 

The organization of the paper is as follows. In the next section we summarize the application of the
Banks-Zaks expansion to mesonic bound states in mass perturbed conformal gauge theories.
In section \ref{leadingorder} the leading order expressions
are presented which are rather straightforward and are given only to fix notation and conventions.
Section \ref{corrections} details the (p)NRQCD calculation of the NLO, NNLO and N$^3$LO corrections.
These are used in section \ref{matching} to attempt to match the non-perturbative low $N_f$ and
perturbative high $N_f$ regions. An assumptions is made explicitly, namely that our decay constant to
meson mass ratios are monotonous as a function of $N_f$, allowing the extraction of an estimate of $N_f^*$, the flavor
number where an abrupt change occurs in the ratios.
Finally in section \ref{conclusions} we conclude and provide an outlook for future work.

\section{Banks-Zaks expansion for bound states}
\label{bankszaks}

In the massless case the
theories inside the conformal window are non-trivial interacting conformal gauge theories with some fixed
point $g_*^2$ depending on $N_f$. At least
sufficiently close to $N_f = 33/2$ there is a single relevant $SU(N_f)$-invariant perturbation of this CFT given by the
flavor singlet fermionic mass term. Its anomalous dimension determines the
dependence of all RG invariant dimensionful quantities on the perturbing parameter $m$.
Besides the mass dependence, there is of course dependence on the fixed point coupling (which depends on
$N_f$) and further explicitly on $N_f$. Schematically, a quantity of dimension $k$ can be written as
$m^{k/(1+\gamma)} F(\eps)$ where $\eps = 33/2 - N_f$ and $\gamma$ is the mass anomalous dimension of the
massless theory. A perturbative expansion can then be developed for small $\eps$, combining all $N_f$-dependence.

Depending on the quantity in question, the function $F(\eps)$ can be determined in perturbation
theory through an expansion in $g^2$. The observables in question for us are quantities related to bound
states: mesons with various quantum numbers. A rigorous perturbative treatment of bound states is given
in the non-relativistic effective theory framework (p)NRQCD, which will be our main method.

Inside the conformal window
massive fermions are always heavy in the (p)NRQCD language. Hence the setup corresponds to zero flavors
of light fermions and $N_f$ flavors of heavy fermions in (p)NRQCD terms. Dimensionful quantities, such as
$f_\pi$, $f_\rho$ or $m_\rho$ are then given as being proportional to $m$ and a double series
expansion in $a(\mu) = g^2(\mu)/(16\pi^2)$ with some RG scale $\mu$ and in $1/m^2$ once a choice of
RG scheme has been made. In the perturbed CFT the natural scale is $\mu=m$ which will be our choice as
well. Thus, we will find schematically,
\bea
f_{\pi,\rho} &=& m a^{3/2}(m) ( b_0 + b_1\, a(m) + \ldots ) \nn \\
m_\rho &=& m ( c_0 + c_1\, a(m) + \ldots ) 
\eea
with some coefficients $b_i(N_f)$ and $c_i(N_f)$ which only depend on $N_f$ and where $\ldots$ refer to
higher order as well as non-analytic terms involving $\log(a(m))$. Naturally, in the
massless limit all three quantities are vanishing. But the massless limit can meaningfully can be taken for
the ratios,
\bea
\frac{f_{\pi,\rho}}{m_\rho} = a_*^{3/2} ( d_0 + d_1\, a_* + d_2\, a_*^2 + \ldots )
\eea
where $a(m\to 0) = a_*$ is the fixed point of the massless theory and the coefficients $d_i(N_f)$ again
depend only on $N_f$. The (p)NRQCD calculation will provide all the coefficients above in the $\msbar$
scheme.

Now the fixed point $a_*$ can trivially be expanded in $\eps$ also,
\bea
a_* = \eps\, ( e_0 + e_1\, \eps + \ldots )
\eea
up to 5-loop order 
\cite{Tarasov:1980au, Larin:1993tp, vanRitbergen:1997va, Czakon:2004bu, Baikov:2016tgj, Herzog:2017ohr}
where the corrections do not contain logarithms only higher orders in $\eps$. 
Once all the explicit $N_f$-dependence of the coefficients 
$d_i(N_f)$ is replaced by $N_f = 33/2 - \eps$ we can expand the final result in $\eps$, leading to,
\bea
\frac{f_{\pi,\rho}}{m_\rho} = \eps^{3/2} ( h_0 + h_1\, \eps + h_2\, \eps^2 + \ldots )\;,
\eea
where again $\ldots$ is short hand for both higher orders in $\eps$ as well as powers of $\log\eps$.
The coefficients of the above series are constants and will be our main result up to N$^3$LO order in
(p)NRQCD for $f_\rho / m_\rho$ and up to
NNLO order for $f_\pi / m_\rho$. The order-by-order results for $f_\rho / m_\rho$ are shown in figure \ref{fvmv} 
where the non-perturbative results for $2\leq N_f \leq 10$ are also indicated. Clearly, the
deviation between the NNLO and N$^3$LO approximations is not large down to $N_f = 12$. The same results
for $f_\pi / m_\rho$ are shown in figure \ref{fpmv} but since the N$^3$LO correction is not available we can
not conclude firmly one way or another about the convergence of the perturbative series in this case.

\section{Leading order}
\label{leadingorder}

The perturbative calculation of $f_\pi$ in the NRQCD~\cite{Caswell:1985ui,Bodwin:1994jh} 
and pNRQCD~\cite{Pineda:1997bj, Brambilla:1999xf, Brambilla:2004jw}
formalism is done by first matching the axialvector current of the heavy quark
and antiquark pair to NRQCD operators, and then computing the NRQCD matrix
elements in pNRQCD in terms of the bound-state
wavefunctions~\cite{Beneke:1999qg, Beneke:2013jia}. 

The matching of the decay constant in NRQCD is expressed in terms of NRQCD
operator matrix elements which scale with powers of $v$, 
the velocity of the heavy quark and antiquark inside the bound state. 
In the perturbative case, $v \sim g^2$, 
so that in order to obtain expressions at NNLO accuracy, it suffices to keep 
corrections up to relative order $v^2$. 
To relative order $v^2$, $f_\pi$ can be written as~\cite{Braaten:1995ej}
\bea
f_\pi &=& \frac{1}{\sqrt{m_\pi}} 
\bigg( c_p \langle 0 | \chi^\dag \psi | \pi \rangle 
\nn \\ && \hspace{8ex} 
- \frac{d_p}{2 m^2} \langle 0 | \chi^\dag (-\tfrac{i}{2} \overleftrightarrow{\bf
D})^2 \psi | \pi \rangle \bigg), 
\eea
where $\psi$ and $\chi^\dag$ are operators that annihilate a heavy quark
and antiquark, respectively, ${\bf D} = \nabla - i g {\bf A}$ is the covariant
derivative, ${\bf A}$ is the gluon field, 
$\chi^\dag \overleftrightarrow{\bf D} \psi 
= \chi^\dag {\bf D} \psi - ({\bf D} \chi)^\dag \psi$, 
$|\pi\rangle$ is the relativistically normalized $\pi$ state at rest, 
and 
$c_p = 1+O(a)$ and $d_p = 1+ O(a)$ are the matching coefficients that are given
by a series in $a$. 
The NRQCD matrix elements can be computed in pNRQCD in terms of the bound-state
wavefunction $\psi(r)$ and the binding energy $E$, which satisfy the
Schr\"odinger equation 
\bea
\left( - \frac{{\nabla}^2}{m} + V(r) \right) \psi(r) = E \psi(r),
\eea
where the potential $V(r)$ is obtained by perturbatively matching pNRQCD to
NRQCD. 
The mass of the bound state is given in terms of the binding energy $E$ by 
\bea
m_\pi &=& 2 m + E,
\eea
and the matrix elements are given by 
\bea
\langle 0 | \chi^\dag \psi | \pi \rangle &=& 
\sqrt{2 N_c} |\psi(0)|, 
\\
\langle 0 | \chi^\dag (-\tfrac{i}{2} \overleftrightarrow{\bf
D})^2 \psi | \pi \rangle 
&=& \sqrt{2 N_c} |\psi(0)| m E, \nn
\eea
where $N_c =3$ is the number of colors. 
These lead to the following expression for $f_\pi$ 
\bea
f_\pi =
\sqrt{\frac{N_c}{m}} \left[ c_p - \left( \frac{c_p}{4} + \frac{d_p}{2}
\right) \frac{E}{m}
\right] |\psi(0)|,
\eea
which is valid up to corrections of order $a^4$. 

Analogously, $f_\rho$ is given in NRQCD by~\cite{Barbieri:1975ki, 
Celmaster:1978yz, Keung:1982jb, Beneke:1999qg}
\bea
f_\rho &=& \frac{1}{\sqrt{m_\rho}}
\bigg( c_v 
\langle 0 | \chi^\dag {\bf \epsilon} \cdot {\bf \sigma} \psi | \rho \rangle
\nn \\ && \hspace{8ex} 
- \frac{d_v}{6 m^2} \langle 0 | \chi^\dag 
{\bf \epsilon} \cdot {\bf \sigma} 
(-\tfrac{i}{2} \overleftrightarrow{\bf
D})^2 \psi | \rho \rangle \bigg),
\eea
where $\sigma$ is a Pauli matrix, $\epsilon$ is the polarization vector for the
$\rho$, and $|\rho\rangle$ is the relativistically normalized $\rho$ state at
rest. Similarly to the case of $f_\pi$, the NRQCD matrix elements appearing in
the expression for $f_\rho$ can be computed in pNRQCD, which lead to the
following expressions 
\bea
\langle 0 | \chi^\dag {\bf \epsilon} \cdot {\bf \sigma} \psi | \rho \rangle &=&
\sqrt{2 N_c} |\psi(0)|,
\\
\langle 0 | \chi^\dag {\bf \epsilon} \cdot {\bf \sigma} 
(-\tfrac{i}{2} \overleftrightarrow{\bf D})^2 \psi | \rho \rangle
&=& \sqrt{2 N_c} |\psi(0)| m E, \nn
\eea
and $m_\rho = 2 m+ E$. From these we obtain 
\bea
f_\rho &=&
\sqrt{\frac{N_c}{m}} \left[ c_v - \left( \frac{c_v}{4} + \frac{d_v}{6} \right)
\frac{E}{m} \right] |\psi(0)|,
\eea
which is again valid up to corrections of order $a^4$. 

At leading order in $a$, it suffices to solve the Schr\"odinger equation 
for the Coulomb potential 
$V(r) = - 4 \pi a C_F/r$, where $C_F = (N_c^2-1)/(2 N_c) = 4/3$. 
In this case the bound-state solutions are known exactly and we obtain for the 
ground state 
\bea
\psi(0) &=& (4 m \pi a C_F)^{3/2}/(8 \pi)^{1/2} [ 1+O(a)],\\ 
E &=& - m (4 \pi a C_F)^2/4 + O(a^3), \nn
\eea
for both the spin-triplet and spin-singlet states. 
From these we obtain the following expressions for 
$f_\pi$, $f_\rho$, and $m_\rho$ that are valid at leading orders in $a$. 
\bea
f_\pi &=&
\sqrt{8 N_c C_F^3} \pi m a^{3/2} [1+O(a)], 
\nn \\
f_\rho &=&
\sqrt{8 N_c C_F^3} \pi m a^{3/2} [1+O(a)], 
\\
m_\rho &=& 2 m [1+O(a^2)]. \nn
\eea
Note that the order-$a$ correction to $m_\rho$ is absent because $E$ begins at
order $a^2$. 

\section{NLO, NNLO and N$^3$LO corrections}
\label{corrections}

Now we discuss the sources of radiative corrections needed to obtain
expressions at NNLO accuracy. 
We first note that 
because $E$ begins at order $a^2$, the leading-order expression for $E$
suffices for obtaining $m_\rho$ at NNLO accuracy. 
The corrections at higher orders in $a$ to NNLO accuracy 
to the decay constants come from the
radiative corrections to the matching coefficients $c_p$
and $c_v$, 
as well as the corrections to the wavefunction at the
origin. 
The corrections to $c_p$ have been computed analytically 
in~\cite{Braaten:1995ej} at NLO and in~\cite{Kniehl:2006qw} at NNLO. 
Likewise, analytical expressions for the 
radiative corrections to $c_v$ are available in~\cite{Barbieri:1975ki,
Celmaster:1978yz} at NLO and in~\cite{Czarnecki:1997vz, Beneke:1997jm} at NNLO. 
As is well known from heavy quarkonium phenomenology, the NNLO corrections to 
$c_v$ and $c_p$ contain logarithms of the NRQCD factorization scale, which must
cancel with the logarithms coming from the renormalization of the NRQCD matrix
elements~\cite{Hoang:2006ty, Chung:2020zqc}. 
The corrections to $|\psi(0)|$ have been computed to NNLO accuracy 
in~\cite{Melnikov:1998ug, Beneke:1999qg} 
for the $S$-wave spin-triplet case. For the spin-singlet case, 
the corrections to $|\psi(0)|$ to NNLO accuracy can be obtained from the results 
in ref.~\cite{Penin:2004ay}. The NNLO corrections contain the logarithms of the
NRQCD factorization scale that cancel against the 
logarithms coming from $c_v$ and $c_p$, so that the decay constants are free of
dependencies on the factorization scale. 
These are sufficient ingredients for computing $f_\pi$ and $f_\rho$ to NNLO 
accuracy.
We note that the dependence on $N_f$ only comes from the matching coefficients,
because all $N_f$ flavors are heavy and are integrated out from the effective
field theory. Also note that a non-vanishing imaginary part of the matching coefficients 
can be discarded at our current level of accuracy. 

Additionally, the N$^3$LO correction to $c_v$ has been computed in
refs.~\cite{Marquard:2014pea, Feng:2022vvk}, 
and the N$^3$LO correction to $|\psi(0)|$ has been computed for the
$S$-wave spin-triplet case in~\cite{Beneke:2014qea, Penin:2014zaa}. 
Together with the NLO correction to $E$ available 
in~\cite{Melnikov:1998ug, Penin:1998kx}
and the NLO correction to $d_v$ available in~\cite{Luke:1997ys, Bodwin:2008vp}, 
these make possible the
computation of $f_\rho$ and $m_\rho$ to N$^3$LO accuracy. 
At N$^3$LO accuracy, in addition to NNLO and N$^3$LO corrections to $c_v$, 
the NLO correction to $d_v$ also contains a logarithm of the
factorization scale, which cancels against the ultrasoft correction to 
$|\psi(0)|$ at N$^3$LO accuracy~\cite{Beneke:2007pj}. 
Because only part of the N$^3$LO correction to $c_v$ is analytically known, 
we only obtain numerical results for the coefficients of the 
order-$a^3$ terms in $f_\rho$. 

We present below the results of the NLO and NNLO corrections for $f_\pi$ and also the 
N$^3$LO correction for $m_\rho$ and $f_\rho$. 

\subsection{$\rho$ mass}
\label{mrhocorrections}

From the binding energy $E$ we have, to N$^3$LO accuracy, 
\bea
\label{mrhocoeffs}
m_\rho &=& c_0 m [1 + c_2 a^2(m) + c_{30} a^3(m) 
\nn \\ && \hspace{4ex} 
+ c_{31} a^3(m) \log a(m) + O(a^4) ]. 
\eea
The order-$a$ term in $m_\rho$ is
zero because $E$ begins at order $a^2$. The first two coefficients
are determined by the leading-order binding energy and to NNLO
and N$^3$LO we obtain the further coefficients,
\bea
\label{mrhocoeffs2}
c_0 &=& 2 \nn \\
c_2 &=& - 2 C_F^2 \pi^2  \\
c_{30} &=& \frac{4}{9} \pi ^2 C_A C_F^2 \left( 66 \log (4 \pi C_F)-97 \right) \nn \\
c_{31} &=& \frac{88}{3} \pi ^2 C_A C_F^2, \nn
\eea
with $C_A = N_c = 3$.

\subsection{$\rho$ decay constant}
\label{frhocorrections}

From the corrections to $c_v$ and $|\psi(0)|$ available to NNLO accuracy,
we obtain
\bea
\label{frhocoeffs}
f_\rho &=& b_0^\rho m a^{3/2}(m) \left( 1 + \right. \\
&& + \left.    \sum_{n=1}^3 \sum_{k=0}^n b^\rho_{nk} a^n(m) \log^{\,k} a(m)
+ O(a^4)
\right). \nn
\eea
The coefficients $b^\rho_{nl}$ up to relative order 
$a^2$ are known analytically and are given by 
\bea
\label{frhocoeffs2}
b_0^\rho &=& \sqrt{8 N_c C_F^3} \pi, \nn \\
b_{10}^\rho &=&
\frac{161}{6}-\frac{11 \pi ^2}{3}+33 \log \left(\frac{3}{16 \pi }\right), 
\\
b_{11}^\rho &=& -33,\nn \\
b_{20}^\rho &=& \left(-\frac{64 \pi^2}{27}+\frac{704}{27}
\right) N_f +\frac{9781 \zeta (3)}{9}
-\frac{27 \pi^4}{8}
\nn \\ && 
+\frac{1126 \pi ^2}{81}
+\frac{9997}{72} 
+\frac{1815 \log ^2\pi}{2}
+\frac{1815}{2} \log ^2\left(\frac{16}{3}\right)
\nn \\ && 
+\log \left(\frac{16}{3}\right) \left(-\frac{2581}{2}+\frac{605 \pi ^2}{3}+1815 \log (\pi )\right)
\nn \\ && 
+\left(\frac{4325 \pi
   ^2}{27}-\frac{2581}{2}\right) \log (\pi )-\frac{256}{81} \pi ^2 \log
(8)
\nn \\ && 
-\frac{1120}{27} \pi ^2 \log \left(\frac{8}{3}\right)-\frac{512}{9} \pi ^2
\log (2),
\nn \\
b_{21}^\rho &=& 
\frac{4325 \pi^2}{27} - \frac{2581}{2} 
+ 1815 \log \left( \frac{16 \pi}{3} \right), 
\nn \\
b_{22}^\rho &=& \frac{1815}{2}. \nn
\eea
The results for the relative order $a^3$ terms are only obtained numerically 
because the analytical result for $c_v$ at N$^3$LO is only
partially known, 
\bea
b^\rho_{30} &=& 0.8198 N_f^2 - 362.7 N_f -1.0901(1) \times 10^6, 
\nn \\
b^\rho_{31} &=& -88.42 N_f -7.7493\times10^5,
\nn \\
b^\rho_{32} &=& -2.1651 \times 10^5,
\nn \\
b^\rho_{33} &=& -2.3292 \times 10^4.
\eea

\subsection{$\pi$ decay constant}
\label{fpicorrections}

From the corrections to $c_p$ and $|\psi(0)|$ available to NNLO accuracy, 
we obtain 
\bea
\label{fpicoeffs}
f_\pi &=& b_0^\pi m a^{3/2}(m) \left(1 + \right. \\
    && \left.    \sum_{n=1}^2 \sum_{k=0}^n b^\pi_{nk} a^n(m) \log^{\,k} a(m)
+ O(a^3)
\right),\nn
\eea
and the coefficients $b^\pi_{nk}$ are given by 
\bea
\label{fpicoeffs2}
b_0^\pi &=& \sqrt{8 N_c C_F^3} \pi, \nn \\
b_{10}^\pi &=& 
\frac{59}{2}-\frac{11 \pi ^2}{3}+33 \log \left(\frac{3}{16 \pi }\right), \\
b_{11}^\pi &=& -33, \nn \\
b_{20}^\pi &=& N_f \left(-\frac{32 \pi ^2}{9}+\frac{344}{9}\right)+
961 \zeta (3)-\frac{27 \pi ^4}{8}
\nn \\ &&
+\frac{1310 \pi^2}{27}+\frac{23053}{72}+\frac{1815 \log ^2 \pi}{2}
+\frac{1815}{2} \log ^2\left(\frac{16}{3}\right)
\nn \\ &&
+\log
   \left(\frac{16}{3}\right) \left(-\frac{2757}{2}+\frac{1271 \pi ^2}{9}+1815 \log \pi \right)
\nn \\ &&
+\left(\frac{1271 \pi^2}{9}-\frac{2757}{2}\right) \log \pi
-\frac{272}{9} \pi ^2 \log 2, \nn\\
b_{21}^\pi &=& 
\frac{1271\pi^2 }{9} - \frac{2757}{2} + 
\frac{1815}{2} \log \left(\frac{256 \pi^2}{9}\right), \nn\\
b_{22}^\pi &=& \frac{1815}{2}. \nn
\eea
Unfortunately at present the N$^3$LO corrections for $f_\pi$ are not available.

\subsection{Banks-Zaks expansion of ratios}

Now that all three quantities of interest are available in perturbation theory
we may expand the ratios in $\eps = 33/2 - N_f$ as
outlined in section \ref{bankszaks}.

Using the 5-loop $\beta$-function 
\cite{Tarasov:1980au, Larin:1993tp, vanRitbergen:1997va, Czakon:2004bu, Baikov:2016tgj, Herzog:2017ohr}
for the expansion of $a_*$ and the perturbative series (\ref{mrhocoeffs}), (\ref{frhocoeffs}) and
(\ref{fpicoeffs}) we obtain the two meson decay constant to mass ratios in numerical form as,
\bea
\label{result1}
\frac{f_\rho}{m_\rho} &=& \eps^{3/2} C_0 
\left( 1 + \sum_{n=1}^3 \sum_{k=0}^{n} C_{nk} \eps^n \log^k\eps + O(\eps^4) \right) \nn \\
\label{result2}
\frac{f_\pi}{m_\rho} &=& \eps^{3/2} C_0 
\left( 1 + \sum_{n=1}^2 \sum_{k=0}^{n} D_{nk} \eps^n \log^k\eps + O(\eps^3) \right)\;, \nn
\eea
with the coefficients,
\bea
\label{cd}
C_0 &=& 0.005826678 \nn \\ \nn \\
C_{10} &=& 0.4487893 \nn \\
C_{11} &=& - 0.2056075 \nn \\
C_{20} &=& 0.2444502\nn \\
C_{21} &=& -0.1624891\nn \\
C_{22} &=& 0.03522870 \\
C_{30} &=& 0.10604(3) \nn\\
C_{31} &=& -0.1128420\nn \\
C_{32} &=& 0.03695458\nn \\
C_{33} &=& -0.005633665\nn
\eea
\bea
\label{cd2}
D_{10} &=& 0.4654041 \nn \\
D_{11} &=& -0.2056075 \nn \\
D_{20} &=& 0.2845697 \\
D_{21} &=& -0.1737620 \nn \\
D_{22} &=& 0.03528692\;. \nn 
\eea
Even though the coefficients (\ref{mrhocoeffs2}), (\ref{frhocoeffs2}) and (\ref{fpicoeffs2}) are dangerously increasing 
in the series (\ref{mrhocoeffs}), (\ref{frhocoeffs}) and (\ref{fpicoeffs}),
the above coefficients of the ratios are much better behaved. This will be important for the
reliability and robustness of our findings.

The coefficients (\ref{cd}) and (\ref{cd2}) are the main results of this paper.

\section{Matching across the conformal window}
\label{matching}

The perturbative calculations are valid close to the upper end of the conformal window
where $\eps = 33/2 - N_f$ is small. Non-perturbative results are available in the 
low $N_f$ region, specifically for $2 \leq N_f \leq 10$, all extrapolated to the chiral and continuum
limit.

With the perturbative results for $f_\rho/m_\rho$ and $f_\pi/m_\rho$
up to N$^3$LO and NNLO order, respectively, at hand we attempt to match them to the non-perturbative
ones. The latter shows that below the conformal window both of our
ratios are constants as a function of $N_f$ to high precision. 
At $N_f = 33/2$ both ratios are vanishing,
and it is natural to expect that both reach zero in a monotonous fashion. Assuming it is indeed the
case we may attempt to interpolate.

\subsection{$f_\rho / m_\rho$}

\begin{figure}
\begin{center}
\includegraphics{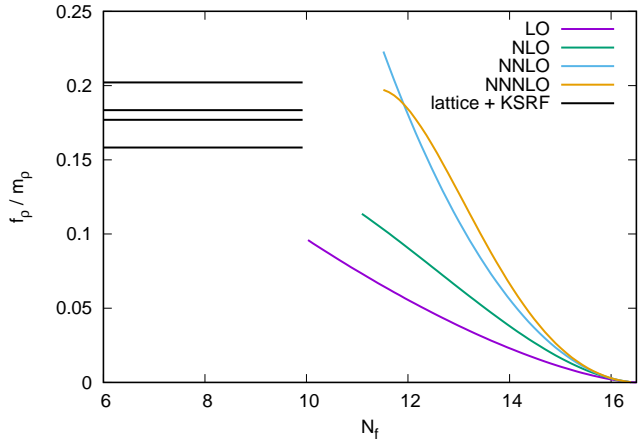}
\caption{The $f_\rho / m_\rho$ ratio in increasing perturbative order as obtained from the Banks-Zaks
expansion in $\eps = 33/2 - N_f$. The non-perturbative result from combined lattice calculations
\cite{Nogradi:2019iek, Nogradi:2019auv, Kotov:2021mgp}
and the KSRF-relation is also shown. The smaller error band corresponds to the uncertainty of the lattice
calculation, the wider one combines this with a conservative estimate of the uncertainty of the
KSRF-relation itself.}
\label{fvmv}
\end{center}
\end{figure}

In order to match our perturbative $f_\rho / m_\rho$ results to the non-perturbative (low $N_f$) region,
continuum and chirally extrapolated lattice results for $f_\rho$ would be needed. These are not available
at the moment, but they are
\cite{Nogradi:2019iek, Nogradi:2019auv, Kotov:2021mgp} for $f_\pi$ in the range $2\leq N_f \leq 10$ 
and the KSRF-relations 
\cite{Kawarabayashi:1966kd, Riazuddin:1966sw} can be used to relate $f_\pi$ and $f_\rho$.
The relation we need is simply $f_\rho = \sqrt{2} f_\pi$.
One does not expect this relation to hold exactly, but even in QCD at finite quark masses it holds to
about 12\% and towards the chiral limit it is expected to hold to even
higher precision. We conservatively assign the 12\% uncertainty as the inherent uncertainty of the
KSRF-relation throughout. Note that in supersymmetric QCD the KSRF-relations have actually been rigorously 
derived \cite{Komargodski:2010mc}.

Hence by combining the non-perturbative lattice results and the KSRF relations we will have access to
$f_\rho / m_\rho$ for $2 \leq N_f \leq 10$. This is shown, together with the Banks-Zaks expansion
(\ref{result1}) order-by-order in figure \ref{fvmv}. The smaller error band displayed for the
non-perturbative result corresponds to the uncertainty of the lattice result while the wider one combines 
this with the estimated 12\% error of the KSRF-relation itself. The dominant uncertainty is from the
latter.
Clearly, the deviation between the NNLO and N$^3$LO results for
$N_f \geq 12$ is not substantial. And curiously, close to $N_f = 12$ the perturbative result reaches the
non-perturbative one almost exactly. More quantitatively, in the range $11.9 \leq N_f \leq 12.1$, the
deviation between the NNLO and N$^3$LO results is at most 4\%, or in the range $11.5 \leq N_f \leq 12.5$
at most 13\%. Hence we conclude that in the region of interest, $N_f \sim 12$, the N$^3$LO result is
robust and reliable.

Assuming $f_\rho / m_\rho$ is a monotonous function of $N_f$ and that around $N_f \sim 12$ the
perturbative result is indeed reliable we are led to conclude that the combination of non-perturbative
and perturbative results cover the entire $N_f$ range. And at twelve flavors an abrupt change occurs in
the ratio which is tempting to identify with the lower end of the conformal window. 
Concretely, we obtain $N_f^* = 12.00(4)$ and $N_f^* = 12.08(6)$ from the NNLO and N$^3$LO
approximations, respectively, if only the uncertainty of the lattice calculation is taken into account. 
If the estimated much larger uncertainty of the KSRF-relation is also taken into account we obtain
$12.0(3)$ and $12.0(5)$ from the NNLO and N$^3$LO approximations, respectively.
Clearly, the NNLO and N$^3$LO approximations agree and lead to $N_f^* = 12$ for integer flavor
numbers. Our line of reasoning cannot of course determine where exactly the twelve flavor theory 
lies, whether 
\cite{Appelquist:2007hu,Appelquist:2009ty,Fodor:2009wk,Fodor:2011tu,Hasenfratz:2011xn,Aoki:2012eq,Fodor:2016zil,Hasenfratz:2016dou,Fodor:2017gtj,Hasenfratz:2017qyr}
it is just below the conformal window and is hence spontaneously broken or just inside and is hence
conformal.

\subsection{$f_\pi / m_\rho$}

\begin{figure}
\begin{center}
\includegraphics{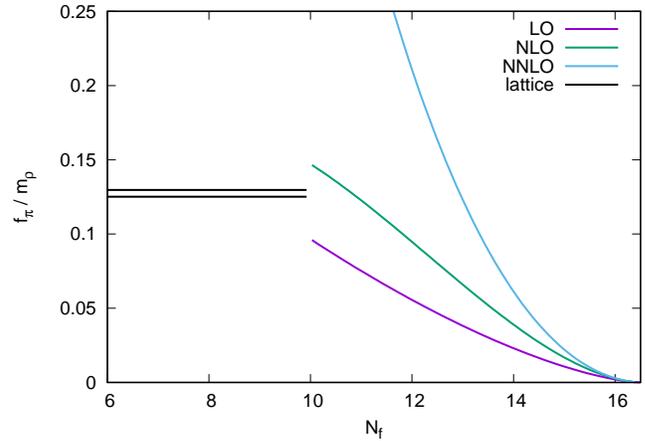}
\caption{The $f_\pi / m_\rho$ ratio in increasing perturbative order as obtained from the Banks-Zaks
expansion in $\eps = 33/2 - N_f$. The non-perturbative result from lattice calculations 
\cite{Nogradi:2019iek, Nogradi:2019auv,Kotov:2021mgp} is also shown.}
\label{fpmv}
\end{center}
\end{figure}

A similar analysis can be performed for $f_\pi / m_\rho$ as well. Here non-perturbative lattice results
are available directly without reliance on any further input. The perturbative calculation could
unfortunately be only carried out to NNLO order though. 
The increasing perturbative orders are shown in figure \ref{fpmv} which also shows the non-perturbative
result obtained from continuum and chirally extrapolated lattice calculations.

The N$^3$LO correction for $f_\rho / m_\rho$ was essential to establish the reliability of the
perturbative series hence we can not make a similar statement for $f_\pi / m_\rho$. We may however
estimate the size of the N$^3$LO correction by assuming that relative to the NNLO result it is comparable
to the case of $f_\rho / m_\rho$. Assuming this is the case we obtain a very similar picture; the
perturbative series seems reliable down to $N_f^* = 13$ where it matches the non-perturbative result.
The only difference relative to $f_\rho / m_\rho$ is the shift in the estimate of the lower end of the
conformal window, from $N_f^* \simeq 12$ to $N_f^* \simeq 13$. This latter estimate should of course be checked by
a genuine N$^3$LO calculation of $f_\pi$ in the future.

\section{Conclusion and outlook}
\label{conclusions}

In this paper we introduced two quantities we believe are useful proxies for the transition between
chirally broken and conformal gauge theories as the flavor number is varied. A minimal requirement for
any such quantity is that it should be well-defined and calculable in the massless limit 
both outside and inside the conformal
window. Outside the conformal window lattice calculations offer a way to obtain results whereas close to
the upper end perturbative ones. Our quantities are related to bound states defined in the mass perturbed
models and the chiral limit is meaningful for both ratios. 

It appears the bridge between the low $N_f$ non-perturbative and high $N_f$ perturbative regions may
not be as large as one might have expected. Current lattice results are available up to $N_f = 10$ and
the main result from this paper is that at $N_f = 12,\, 13$ the perturbative series might be reliable if
calculations are performed up to N$^3$LO order leaving only the $N_f = 11$ model to be interpolated.
Interestingly, at least for the $f_\rho / m_\rho$ ratio, the perturbative N$^3$LO result at $N_f = 12$ agrees with the
non-perturbative $N_f = 10$ lattice calculation (and the ratio is approximately constant for $2 \leq N_f
\leq 10$). If we assume the ratio is a monotonously decreasing function of $N_f$, which is a natural
assumption based on the behavior at $N_f = 10$ and $N_f = 33/2$, we conclude that a matching between
the low $N_f$ non-perturbative and high $N_f$ perturbative regions is possible with an abrupt change
at $N_f \simeq 12$. It is tempting to identify this with the lower end of the conformal window $N_f^*
\simeq 12$.

Our other ratio, $f_\pi / m_\rho$ offers a similar analysis, but unfortunately at the moment only NNLO
perturbative results are available. The reliability of the perturbative series cannot be judged from the
NLO and NNLO corrections alone, in fact it is clear from the behavior of $f_\rho / m_\rho$ that the N$^3$LO
correction is mandatory in order to conclude. Such a calculation of $f_\pi$ within (p)NRQCD seems feasible and
will be pursued in the future. Meanwhile, we have estimated the relative size of the unknown N$^3$LO correction
for $f_\pi / m_\rho$ from that of $f_\rho / m_\rho$. Assuming that this is justified we are led to
believe that a matching between the non-perturbative and perturbative regions is possible at $N_f \simeq 13$ 
with a similarly abrupt change at this value. Hence the estimate shifted to $N_f^* \simeq 13$, however
it is important to stress that a genuine N$^3$LO calculation of $f_\pi$ should be sought first. 

Needless to say, we have nothing firm to conclude about the $N_f = 12$ model, whether it is just inside
or just outside the conformal window. 

In general it is important to remember a key assumption underlying our entire calculation;
namely that the only $SU(N_f)$-invariant relevant perturbation of the conformal field theories we discuss
is the fermionic mass term. This is certainly correct for small $\eps$ but might not 
hold for a sufficiently strongly coupled CFT, for instance it
is conceivable that a 4-fermi term becomes relevant. Addressing this potential situation is beyond the
scope of the present paper but we hope to return to it in the future.

\section*{Acknowledgment}

We would like to thank George Fleming, Yuchiro Kiyo and Alexander Penin for useful correspondence.
The work of H.S.C. is supported by the National Research Foundation of Korea
(NRF) Grant funded by the Korea government (MSIT) under Contract No.
NRF2020R1A2C3009918 and by a Korea University grant. The work of D.N.
is supported by the NKFIH grant KKP-126769. All authors
contributed equally to this work.

\end{document}